\documentclass[twocolumn,showpacs,preprintnumbers,amsmath,amssymb
 aps,
 prb,
 lengthcheck,%
]{revtex4-1}
\usepackage{amssymb}
\usepackage{graphicx}
\usepackage{dcolumn}
\usepackage{bm}
\usepackage{hyperref}
\usepackage{makeidx}

\usepackage[english]{babel} 
\usepackage[latin1]{inputenc}
\usepackage{color}
\usepackage{graphicx}
\usepackage{amsmath}

\usepackage{latexsym}
\usepackage{amsthm}

\makeindex
\def\>{\right\rangle}
\def\<{\left\langle}
\def\be{\begin{equation}}
\def\ee{\end{equation}}
\def\ba{\begin{array}{l}}
\def\ea{\end{array}}
\def\beq{\begin{eqnarray}}
\def\eeq{\end{eqnarray}}

\begin{document}

 
\title{Non-local interference and Hong-Ou-Mandel collisions\\ of single Bogoliubov quasiparticles}

\author{D. Ferraro$^{1,2}$, J. Rech$^{1}$, T. Jonckheere$^{1}$, and T. Martin$^{1}$ }
\affiliation{$^1$ Aix Marseille Universit\'e, Universit\'e de Toulon, CNRS, CPT, UMR 7332, 13288 Marseille, France\\
$^2$ D\'epartement de Physique Th\'eorique, Universit\'e de Gen\`eve, 24 quai Ernest Ansermet, CH-1211 Geneva, Switzerland}

\date{\today}

\begin{abstract}
We consider a device which allows to create and probe single Majorana fermions, in the form of Bogoliubov quasiparticles. It is composed of two counter-propagating edge channels, each put in proximity with a superconducting region where Andreev reflection operates, and which thus converts electrons into Bogoliubov quasiparticles. The edge channels then meet at a quantum point contact where collisions can be achieved. A voltage biased version of the setup was studied in Phys. Rev. Lett. \textbf{112}, 070604 (2014) and showed non-local interference phenomena and signatures of Bogoliubov quasiparticle collisions in the high frequency noise characteristics at the output, constituting an evidence of the Majorana fermion nature of these excitations. Here, voltage biased leads are replaced by single electron sources in order to achieve collisions of single Bogoliubov quasiparticles, with the major advantage that zero-frequency noise measurements are sufficient to access the intimate nature of Bogoliubov wave-packets. We compute the injection parameters of the source, and go on to investigate the Hanbury-Brown and Twiss and Hong-Ou-Mandel signal at the output, as a function of the mixing angle which controls the electron/hole component of the Bogoliubov wave-packet. In particular, information on the internal structure of the Bogoliubov quasiparticle can be recovered when such a quasiparticle collides with a pure electron. Experimental feasibility with singlet or triplet superconductors is discussed.    
\end{abstract}

\maketitle
\section{Introduction}
Electron quantum optics \cite{Bocquillon13b} offers the unique possibility to apply long standing concepts and ideas developed in the framework of quantum optics, to individual electronic wave-packets propagating in condensed matter systems. Seminal experiments such as the Hanbury-Brown-Twiss (HBT) \cite{Hanbury56} and the Hong-Ou-Mandel (HOM) \cite{Hong87} interferometers have been performed with periodic trains of electrons and holes produced by means of single electron sources based on driven mesoscopic capacitors \cite{Feve07} or properly designed Lorentzian voltage pulses.\cite{Dubois13b} These excitations interfere at the electronic equivalent of a half-silvered mirror -- a quantum point contact (QPC) -- and the outgoing current and noise signals are measured. \cite{Mahe10,Grenier11} The obvious differences between electrons and photons related to the Pauli principle,\cite{Jonckheere12} the presence of the Fermi sea,\cite{Bocquillon12}  and the mutual interaction between the electrons \cite{Bocquillon13, Wahl14} have to be properly taken into account in order to correctly interpret the experimental observations. 

Until now, experiments have been carried out in two dimensional electron gas either in absence of magnetic field \cite{Dubois13b} or by exploiting the ballistic propagation and chirality of integer quantum Hall edge channels.\cite{Bocquillon13b, Bocquillon12, Bocquillon13} Proposals for electron quantum optics experiments have been presented to extend these ideas also to other topological states of matter like two \cite{Inhofer13, Hofer13, Ferraro14b} and three \cite{Khan14} dimensional topological insulators, where new interesting features appear as a consequence of helicity and spin-momentum locking.\cite{Ferraro14b} While experimental/theoretical efforts promise future realizations of such quantum optics scenarios in these newly discovered states of matter, more ``conventional'' condensed matter systems involving superconducting (SC) elements ought to be revisited in view of electron quantum optics applications. This is the goal of the present work.  

\begin{figure}[h]
\centering
\includegraphics[scale=0.4]{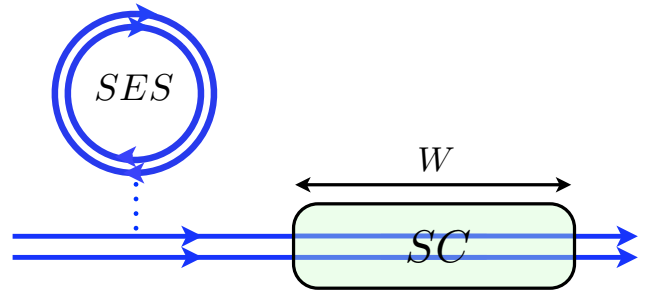}
\caption{(Color online). Schematic view of a single electron source (SES) injecting electron and hole wave-packets into a quantum Hall edge state at filling factor $\nu=2$ coupled with a SC contact of length $W$.}
\label{fig1}
\end{figure}

Indeed, in a seminal work \cite{Beenakker14} a setup in the integer quantum Hall regime was proposed, where continuous flows of electrons are injected in the proximity of SC contacts of finite length before reaching a QPC. Andreev reflection\cite{Andreev64} by the contacts converts electrons into Bogoliubov quasiparticles, namely coherent superpositions of electrons and holes with opposite or equal spin depending on the singlet or triplet nature of the SC coupling. \cite{Beenakker14, Ostaay11}  The outgoing cross-correlated current fluctuations are measured at finite frequency and show a remarkable non-local dependence on the difference in the superconducting phases between two SC contacts as a consequence of the collision of Bogoliubov quasiparticles. As argued in Refs. \onlinecite{Chamon10,Beenakker14}, the fact that Bogoliubov quasiparticle creation operators are related by a unitary transformation to their annihilation counterpart justifies their qualification as Majorana fermions. This constitutes an interesting alternative proposal for Majorana fermion\cite{Majorana37} detection compared to those put forward in topological superconductor devices, which are mainly focused on the investigation of Majorana zero energy modes. \cite{Akhmerov09, Alicea12} In Ref. \onlinecite{Beenakker14}, the annihilation property of Bogoliubov quasiparticle beams at a QPC showed a periodic dependence on the phase difference between the two superconductors: a clear manifestation of a non-local interference effect.

Yet this work focused on electron ``beams'' in the vicinity of SC, and therefore fails to address the single shot creation and collision of two Bogoliubov quasiparticles. In this paper we investigate an analogous setup where voltage electrodes are replaced by single electron sources (see Fig. \ref{fig1}). This allows us to analyze the properties of Bogoliubov excitations at the \emph{single quasiparticle level}, shading light on various aspects which are not explicitly discussed in Ref. \onlinecite{Beenakker14}. 

One of the experimental challenges implied by the diagnosis of Ref. \onlinecite{Beenakker14} is the fact that collisions between Majorana fermions need to be detected through high frequency noise measurements, which currently require an on-chip quantum device\cite{Lesovik_Aguado_Deblock} attached to the sample. Unfortunately, the measurement of the high frequency noise characteristics of a normal metal-superconductor junction as needed here, and which was computed more than a decade ago in Ref. \onlinecite{Torres01}, has so far eluded experimental realization. On the contrary,
with single particle/quasiparticle injection, the requirement for measuring high frequency noise is lifted: zero frequency noise measurements are sufficient to probe physically relevant effects, and they have been already successfully implemented with single electron sources recently.\cite{Mahe10}

Our first aim is to provide information about the nature of the excitations reaching the QPC. We will first characterize the injection process in terms of current and noise, focusing on the role played by the non-conservation of the charge and the conservation of the excitation number which identify the SC. We then move to interferometric configurations with a QPC. In the HBT case only one of the sources of Bogoliubov excitations is ON and measurements return the partition noise associated with the injected quasiparticles, while in the HOM case both sources are ON and measurements can access the intimate structure of these peculiar quasiparticles. This is particularly evident when one of the sources injects a reference state, namely an electron, while the other emits a more general Bogoliubov excitation. In this case the setup can be used to carry out a true spectroscopy of the unknown incoming state, the sign of the outgoing noise depending on the relative weight of the electron and hole component of the superposition. Quite remarkable in this context is the case of a zero charge Bogoliubov excitation. Here the outgoing current as well as the HBT noise contribution are zero, but the HOM noise is maximal as a consequence of the non-trivial structure of the quasiparticle. Finally, it is worth noticing that the considered setup shows no dependence on the SC phase at the level of the averaged current (first order coherence), while oscillations dependent on the SC phase difference appears in the noise (second order coherence).

The paper is divided as follows. In Section \ref{Model} we recover the results of Ref. \onlinecite{Beenakker14} about the action, in terms of a transfer matrix, of a SC contact on an incoming electronic wave-packet originating from an integer quantum Hall system at filling factor $\nu=2$, where spin singlet SC coupling is expected. Section \ref{Current} discusses the current and the excitation density outgoing from the SC contact with particular attention to the role played by the non-conservation of the charge and the conservation of the excitation number associated with the Bogoliubov-De Gennes Hamiltonian describing the system. The fluctuations of the current outgoing from the contact are also investigated in comparison with the results obtained for the conventional single electron source. In Section \ref{Noise} we investigate the cross-correlation noise outgoing from the QPC, described in terms of a scattering matrix. In particular we consider the HBT contribution (\ref{Section_HBT}) where only one source is ON. Next, we deal with the HOM noise (\ref{Section_HOM}) obtained when the two sources are ON. The limitations occurring in extending the same analysis for an integer quantum Hall state at $\nu=1$ with triplet SC coupling are discussed in Section \ref{Triplet}. The Appendix contains analytical evaluations of the most relevant quantities.

\section{Model}\label{Model}

Let us start by discussing the case of two quantum Hall edge channels unresolved in spin, namely the boundary of a Hall bar at filling factor $\nu=2$, in which we neglect the Zeeman splitting and the inter-channel interaction. \cite{Giazotto05} These edge states are coupled to a single electron source (SES) and to a SC contact of length $W$ (see Fig. \ref{fig1}). The SES injects into the channels an electron (hole) with well defined  wave-packets which can have exponential \cite{Feve07} or Lorentzian \cite{Dubois13b} profile in time, depending on the considered experimental set-up. In the following we attempt to remain as general as possible in order to derive expressions which are valid for all the physically relevant cases.  As shown in Ref. \onlinecite{Beenakker14} a spin-singlet coupling between the Hall channel and the SC contact can be realized in graphene and is favored by the small spin-orbit coupling associated with this material.\cite{Popinciuc12, Rickhaus12} 

The action of the SC on incoming electrons with energy below the induced SC gap $\Delta$ can be described in terms of an energy dependent $4\times 4$ transfer matrix $\mathcal{M}$, constrained by unitarity and particle-hole symmetry:\cite{Ostaay11, Beenakker14}

\be
\mathcal{M}(\xi)= \left(\tau_{x}\otimes I\right) \mathcal{M}^{*}(-\xi) \left(\tau_{x}\otimes I \right). 
\label{particle_hole}
\ee
In the above expression  $I$ is the identity in spin space, while from now on we indicate with $\tau_{i}$ ($i=x, y, z$) the Pauli matrices acting on the electron-hole space and with $\sigma_{i}$ ($i=x, y, z$) the ones related to the spin degree of freedom. According to this, the transfer matrix $\mathcal{M}$ is applied to a $4$-component spinor state: \cite{Beenakker14}

\be
c(\xi)= 
\left( 
\begin{matrix}
c_{e, \uparrow}(\xi) \\
c_{e,\downarrow}(\xi)\\
c_{h,\uparrow}(\xi)\\
c_{h, \downarrow}(\xi)\\
\end{matrix}
\right),
\label{spinor}
\ee
 where $e$ ($h$) indicates the electron (hole) state and $\uparrow$ ($\downarrow$) the up (down) spin direction. The particle-hole symmetry in Eq. (\ref{particle_hole}) leads to the constraint: \cite{Alicea12}
 
 \be
 c(\xi)= \tau_{x}\otimes I \left[ c^{\dagger}(-\xi)\right]^{T}.
 \label{particle_hole2}
 \ee
 
The explicit form of this transfer matrix can be deduced starting from the Bogoliubov-De Gennes equation (see Refs. \onlinecite{Ostaay11, Beenakker14} for the details of the derivation) and reads:

\be
\mathcal{M}(\xi)= e^{i \xi \delta} e^{i \gamma \tau_{z}} \mathcal{U}(\alpha, \phi, \beta) e^{i \gamma' \tau_{z}}=e^{i \xi \delta} \tilde{\mathcal{M}}.
\label{transfer}
\ee
This explicitly takes into account the effects of Andreev reflection induced by the SC contacts. In the above expression $\xi$ is the energy of the incoming excitation, $\delta=W/v$ is the time required for the excitation to cross the SC region ($v$ the velocity of propagation along the quantum Hall edge channel), $\alpha\approx W/l_{s}$ ($l_{s}=\hbar v/\Delta$ the proximity-induced coherence length) and $\beta \approx W/l_{m}$ ($l_{m}=(\hbar /e B)^{\frac{1}{2}}$ the magnetic length of the Hall system, with $B$ the applied perpendicular magnetic field). By comparing the expression for the upper critical magnetic field in a Type II SC in terms of the coherence length $B_{c}= \Phi_{0}/(2 \pi l^{2}_{s})$ and the definition of magnetic length $B=  \Phi_{0}/(2 \pi l^{2}_{m})$ ($\Phi_{0}$ the elementary flux quantum) one finds that, in order not to destroy the SC, the condition $l_{s}\ll l_{m}$, and consequently $\alpha \gg \beta$, is required in this case. The parameters $\gamma, \gamma'$ take into account the relative phase shifts of electrons and holes in presence of the magnetic field \cite{Hoppe00, Ostaay11, Beenakker14} and with $\phi$ we indicate the order parameter phase of the SC state.

The unitary matrix $\mathcal{U}$ which appears in Eq. (\ref{transfer}) is given by: \cite{Beenakker14} 
\be
\mathcal{U}(\alpha, \phi, \beta) =\exp{\left[i \alpha \sigma_{y} \otimes (\tau_{x} \cos{\phi}+ \tau_{y} \sin{\phi})+ i\beta \tau_{z} \right]}.
\ee
In order to further simplify the expression for $\mathcal{M}$, it is possible to eliminate the $\tau_{z}$ term from $\mathcal{U}$ using the Baker-Hausdorff relation. We then obtain:

\be
\mathcal{M}(\xi)= e^{i \xi \delta} e^{i \Gamma \tau_{z}} \mathcal{U}(\tilde{\theta}, \phi, 0) e^{i \Gamma' \tau_{z}}
\label{transfer_exp}
\ee
where we introduced the new angle $\tilde{\theta}$ such that: 

\beq
\sin{\tilde{\theta}}&=& \frac{\alpha}{\theta} \sin{\theta}
\label{sin_theta}\\
\cos{\tilde{\theta}}&=& \sqrt{\cos^{2}{\theta}+ \frac{\beta^{2}}{\theta^{2}} \sin^{2}{\theta}},
\eeq
with $\theta= \sqrt{\alpha^{2}+\beta^{2}}$. The new phase shifts are given by: 

\beq
\Gamma= \gamma+ \Omega,\\
\Gamma'= \gamma'+ \Omega
\eeq
with $\Omega= \arctan\left(\frac{\beta}{\theta} \tan{\theta} \right)$.  

The unitary matrix $\mathcal{U}$ can then be rewritten in the simpler form 

\be
\mathcal{U}(\tilde{\theta}, \phi, 0) =\left( 
\begin{matrix}
\cos{\tilde{\theta}}&0 & 0&  e^{-i \phi} \sin{\tilde{\theta}}\\
0&\cos{\tilde{\theta}} & -e^{-i \phi} \sin{\tilde{\theta}}&0\\
0&e^{i \phi} \sin{\tilde{\theta}} & \cos{\tilde{\theta}}&0\\
-e^{i \phi} \sin{\tilde{\theta}} &0&0& \cos{\tilde{\theta}}
\end{matrix}
\right)
\label{S_matrix_tilde}
\ee 
in terms of the new variable $\tilde{\theta}$.

We consider now a SES injecting a spin up electron into the SC region:\cite{Feve07, Note1}

\beq
|\varphi \rangle&=& \int^{+\infty}_{-\infty}d \tau \varphi_{e}(\tau) \Psi^{\dagger}_{\uparrow}(\tau) |F\rangle \nonumber\\
&=& \frac{1}{\sqrt{2 \pi}} \int^{+\infty}_{-\infty} d \xi \tilde{\varphi}_{e}(\xi) c^{\dagger}_{e, \uparrow}(\xi) |F\rangle,
\label{wave_packet} 
\eeq
with $\varphi_{e}(\tau)$ $(\tilde{\varphi}_{e}(\xi))$ a normalized wave-packet in the time (energy) domain well localized above the Fermi sea $|F\rangle$. The chemical potential of the Fermi sea will be considered as the reference for measuring the energy and we assume the zero temperature limit where the Fermi distribution is given by $f_{e}(\xi)=f_{h}(\xi)=\Theta(-\xi)$. Analogous expressions can be considered for the other possible incoming states (spin down electrons, spin up and spin down holes). 

Because of the action of the transfer matrix, the corresponding state outgoing from the SC is a Bogoliubov quasiparticle given by the coherent superposition of one electron and one hole with opposite spin outgoing from the SC region:
\beq
|\mathcal{B}\rangle&=&\mathcal{W}_{e} |e, \uparrow \rangle+ \mathcal{W}_{h} |h, \downarrow \rangle\nonumber\\
&=&\cos{\tilde{\theta}} |e, \uparrow \rangle+ \sin{\tilde{\theta}}e^{-i\Phi} |h, \downarrow \rangle
\label{Bogoliubov}
\eeq
with $\Phi=2 \Gamma- \phi$ and $|e, \uparrow \rangle$, $|h, \downarrow \rangle$ a short notation for the electron and hole outgoing states respectively.

\section{Current and particle density} \label{Current}

We consider the expressions for the averaged total current and particle density outgoing from the considered device. In the following we report only the results, the detailed derivation being developed in Appendix \ref{AppA}. These quantities are given by the sum of spin up and spin down contributions:
\beq
\langle \varphi |I(t) |\varphi \rangle &=&-e v\langle \varphi |:\Psi^{\dagger}(t) \tau_{z} \Psi(t): |\varphi 
\label{current_def}\rangle\\
\langle \varphi |\rho (t) |\varphi \rangle &=& v\langle \varphi |:\Psi^{\dagger}(t) \Psi(t): |\varphi \rangle
\label{density}
\eeq
where the notation $:...:$ corresponds to the usual normal ordering with respect to the Fermi sea and where $-e$ ($e>0$) is the electron charge. Note that, in the above expressions, the definition 
\be
\Psi(t)=\frac{1}{\sqrt{4\pi}} \int^{+\infty}_{-\infty} d \xi e^{-i \xi t} \mathcal{M}(\xi) c(\xi) 
\ee
for the field operator outgoing from the superconducting region is required to avoid double counting associated with the particle-hole symmetry. \cite{Beenakker14}

Applying Wick's theorem and considering well localized wave-packets in the positive energy domain, the current reduces to: 
\beq
\langle \varphi |I(t) |\varphi \rangle&=&-e \mathrm{Tr} \left( \mathcal{P}_{s} \tilde{\mathcal{M}}^{\dagger} \tau_{z} \tilde{\mathcal{M}} \mathcal{P}_{s}\right) \varphi(t- \delta) \varphi^{*}(t- \delta)\nonumber \\
&=&-e\cos(2\tilde{\theta}) \varphi(t- \delta) \varphi^{*}(t- \delta),
\label{current}
\eeq
where we introduced the projector:
\be
\mathcal{P}_{s}= \left(\frac{1+\tau_{z}}{2}\right)\otimes \left(\frac{1+\sigma_{z}}{2}\right)
\label{projector}
\ee
in order to properly take into account the injection of an individual spin up electron.

 The physical meaning of this result clearly emerges by recalling Eq. (\ref{Bogoliubov}). The outgoing electronic current of Eq. (\ref{current}) differs from the incoming one \cite{Grenier11, Ferraro13} $\langle \varphi |I_{in}(t) |\varphi \rangle \equiv -e \varphi(t) \varphi^{*}(t)$, by a time delay $\delta$ and by a factor $\cos(2\tilde{\theta})$, which takes into account the effect of the SC region when converting electrons into holes through Andreev reflections. This factor is simply given by the difference between the probability $P_{e}=|\mathcal{W}_{e}|^{2}= \cos^{2}{\tilde{\theta}}$ for the incoming electron to emerge as an electron and $P_{h}=|\mathcal{W}_{h}|^{2} =\sin^{2}{\tilde{\theta}}$ to be converted into a hole. For $\tilde{\theta}=0$ ($P_{e}=1$ and $P_{h}=0$) the SC contact only induces a delay, while for $\tilde{\theta}=\pi/2$ ($P_{e}=0$ and $P_{h}=1$) the incoming electron is completely converted into a hole and a Cooper pair enters into the SC. More importantly, for $\tilde{\theta}=\pi/4$ ($P_{e}=P_{h}=1/2$) the electron and hole contributions compensate and no averaged current flows out of the device. As will be clear in the following this zero averaged current still shows fluctuations which can be detected via noise measurements. 

According to the above discussion the charge outgoing from the SC contact:  

\be
\mathcal{Q}= \int^{+\infty}_{-\infty} dt \langle \varphi |I(t) |\varphi \rangle= -e \cos(2\tilde{\theta})
\label{charge}
\ee
is not conserved as a consequence of the creation/destruction of Cooper pairs in the SC. Conversely, according to the unitarity of the scattering matrix, the outgoing excitation density is given by: 
\beq
\langle \varphi |\rho(t) |\varphi \rangle= \varphi (t- \delta) \varphi^{*}(t- \delta)= \langle \varphi |\rho_{in}(t-\delta) |\varphi \rangle,\nonumber\\
\eeq
which implies a mere time delay $\delta$ with respect to the incoming one. In this case the prefactor is given by $P_{e}+P_{h}=1$ as a consequence of the fact that the incoming electron can only emerge from the device as an electron or as a hole and cannot be absorbed by the SC.  This illustrates the conservation of the number of injected excitations: 
\be
\mathcal{N}= \int^{+\infty}_{-\infty} d t \langle \varphi |\rho(t) |\varphi \rangle=\int^{+\infty}_{-\infty} dt \langle \varphi |\rho_{in}(t-\delta) |\varphi \rangle=1
\ee
as expected. Note that both the non-conservation of the charge and the conservation of the excitation number are encoded in the Bogoliubov-De Gennes Hamiltonian which is at the origin of the transfer matrix of Eq. (\ref{transfer}) (see Ref. \onlinecite{Ostaay11}).

Proceeding along the same way, we consider the current fluctuations at the output of the SC contact. In the zero frequency limit they are given by: \cite{Note2}
\beq
S_{source}&=&\int^{+\infty}_{-\infty} dt dt' \left[\langle \varphi| I(t) I(t')  |\varphi\rangle_{c}\right]\nonumber\\
&=& e^{2} \left\{1- \left[\mathrm{Tr} \left( \mathcal{P}_{s} \tilde{\mathcal{M}}^{\dagger} \tau_{z} \tilde{\mathcal{M}} \mathcal{P}_{s}\right)\right]^{2}\right\}\\
&=& e^{2}\sin^{2}(2\tilde{\theta})
\label{noise_out}
\eeq
where $\langle \varphi |A B|\varphi \rangle_{c}= \langle \varphi| A B |\varphi \rangle-\langle \varphi| A |\varphi \rangle\langle \varphi | B |\varphi \rangle$ is the connected two point correlator of generic operators $A$ and $B$. 

It is worth noticing that the above quantity is proportional to the product $P_{e} \cdot P_{h}$. Therefore it is zero in absence of a SC contact ($\tilde{\theta}=0$), as expected, \cite{Mahe10, Bocquillon13b} and also for $\tilde{\theta}=\pi/2$ when the incoming electron is completely converted into a hole. Even more interestingly is the fact that it reaches its maximum for $\tilde{\theta}=\pi/4$, when the outgoing averaged current is zero due to the action of the SC contact. 

\section{Cross-correlated noise in a QPC geometry}\label{Noise}

\begin{figure}[h]
\centering
\includegraphics[scale=0.28]{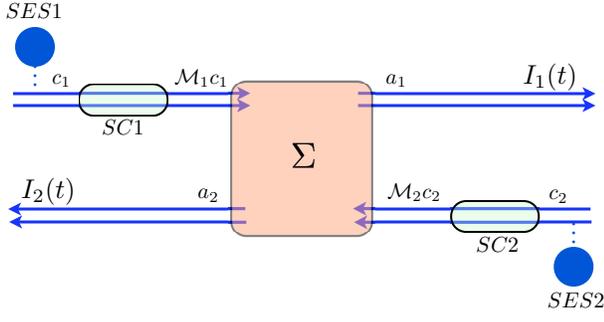}
\caption{(Color online). QPC geometry for Bogoliubov quasiparticles. Two SES ($SES1$ and $SES2$) inject electrons, described by the operators $c_{j}$ ($j=1,2$), into two SC contacts ($SC1$ and $SC2$). The outgoing excitations, described by $\mathcal{M}_{j}c_{j}$, reach the QPC and are partitioned according to the scattering matrix $\Sigma$. The outgoing currents $I_{j}(t)$ (written in terms of the operators $a_{j}$) are then recollected in order to access the cross-correlated noise.}
\label{fig2}
\end{figure}

Once the device is characterized as an emitter of individual Bogoliubov quasiparticles we need to investigate the outgoing cross-correlated noise in a QPC geometry where one or two SES ($SES1$ and $SES2$) inject electronic wave-packets with spin up in the vicinity of one or two SC regions (see Fig. \ref{fig2}). 

The scattering matrix associated with the QPC reads: \cite{Grenier11, Beenakker14}
 \be
\Sigma=
\left( 
\begin{matrix}
\sqrt{1-R} & - \sqrt{R}\\
\sqrt{R} & \sqrt{1-R}\\
\end{matrix}
\right)
\label{Sigma_matrix}
\ee
where $R$ is the probability of reflection. Consequently the annihilation spinors outgoing from the QPC are related to the ones emitted by the two SES through the equations:
\beq
a_{1}&=& \sqrt{1-R} \mathcal{M}_{1}c_{1} -\sqrt{R} \mathcal{M}_{2} c_{2,} \\
a_{2}&=& \sqrt{R} \mathcal{M}_{1}c_{1} + \sqrt{1-R} \mathcal{M}_{2} c_{2},
\eeq
where the energy dependence has been omitted for notational convenience.

The zero frequency cross-correlated noise outgoing from the QPC is given, in the wave-packet approximation, by: 

\be
S=\int^{+\infty}_{-\infty} d t d t' \langle \varphi | I_{1}(t) I_{2}(t') |\varphi \rangle_{c}.
\ee
By properly taking into account the particle-hole symmetry in Eq. (\ref{particle_hole}) it is possible to recast the above expression in the form: 

\be
S= e^{2} v^{2} \int^{+\infty}_{0} d \xi d \rho \langle \varphi| a^{\dagger}_{1}(\xi) \tau_{z} a_{1}(\xi)  a^{\dagger}_{2}(\rho) \tau_{z} a_{2}(\rho) |\varphi\rangle_{c}
\ee
where the integrals run over positive energies only. This will lead to important simplification in the following when discussing the HBT and the HOM interferometers in detail.

\subsection{Hanbury-Brown-Twiss contribution}\label{Section_HBT}

When only one of the two SES (indicated for sake of generality with $j$) is ON we obtain the HBT contribution to the noise. Here, in the zero temperature limit, the injected excitations crossing the SC contact are converted into Bogoliubov quasiparticles which reach the QPC and get partitioned.\cite{Bocquillon12} The expression for this contribution to the noise is:

\beq 
S^{HBT}_{j}&=&-e^{2} R(1-R)  \left[\mathrm{Tr}\left( \mathcal{P}_{s} \tilde{\mathcal{M}}_{j}^{\dagger}\tau_{z}  \tilde{\mathcal{M}}_{j} \mathcal{P}_{s}\right)\right]^{2} \\
&=&  -e^{2} R(1-R) \cos^{2}(2\tilde{\theta}_{j}).
\label{HBT}
\eeq
This represents the shot noise associated with a wave-packet carrying charge $\mathcal{Q}$ (see Eq. (\ref{charge})) and is therefore proportional to $\mathcal{Q}^{2}$. In absence of SC ($\tilde{\theta}=0$), the transfer matrix reduces to the identity ($\mathcal{M}_{j}(\xi)= I\otimes I$) and the above expression becomes: 
\be
S^{HBT}=-e^{2} R(1-R)
\label{No_SC}
\ee
as expected.\cite{Bocquillon13b, Ferraro13} Note that for $\tilde{\theta}_{j}=\pi/4$, the state which reaches the QPC, given by a balanced coherent superposition of electrons and holes, generates no noise at all because it can be seen as an individual excitation bearing zero charge. For $\tilde{\theta}_{j}=\pi/2$, the electron is completely converted into a hole when put in contact with the superconductor, and we recover again the result of Eq. (\ref{No_SC}).   

\subsection{Hong-Ou-Mandel contribution}\label{Section_HOM}
If both the SES are ON we obtain the HOM noise signal \cite{Jonckheere12, Bocquillon13, Bocquillon13b}

\be
S^{HOM}= \Delta S^{HOM}+ S^{HBT}_{1}+S^{HBT}_{2}
\ee
with
\beq
&&\Delta S^{HOM}/S_{0}= \nonumber\\
&&2 A(\delta_{1}- \delta_{2}-\eta) \mathrm{Tr} \left[ \mathcal{P}_{s} \tilde{\mathcal{M}}_{1}^{\dagger}\tau_{z}  \tilde{\mathcal{M}}_{2} \mathcal{P}_{s}\right] \mathrm{Tr} \left[ \mathcal{P}_{s} \tilde{\mathcal{M}}_{2}^{\dagger}\tau_{z}  \tilde{\mathcal{M}}_{1} \mathcal{P}_{s}\right] = \nonumber\\
&&A(\delta_{1}-\delta_{2}-\eta)\times \nonumber\\
&&\left[ 1+ \cos(2\tilde{\theta}_{1})\cos(2\tilde{\theta}_{2})-\cos(\Phi_{12}) \sin(2\tilde{\theta}_{1})\sin(2 \tilde{\theta}_{2})\right]
\label{HOM_general}
\eeq
being $S_{0}= e^{2} R(1-R)$.
This constitutes a central analytical result of this work, as it addresses the HOM collision of two arbitrary Bogoliubov quasiparticles.
In the above expression $\eta$ is the time delay in the emission between the two SES, $A(\tau)= |\mathcal{A}(\tau)|^2$ with:

\be
\mathcal{A}(\tau)=\int^{+\infty}_{-\infty} dt  \varphi^{*}(t- \tau) \varphi(t)
\ee
the overlap between identical wave-packets emitted with a delay $\tau$ and: 

\be
\Phi_{jk}= 2 \Gamma_{j}- 2\Gamma_{k}-\phi_{j}+\phi_{k}
\ee
a phase which is reminiscent of the one appearing in Eq. (\ref{Bogoliubov}). 

When the two SC regions only differ in their order parameter phase and the two SES are properly synchronized ($\eta=0$) one obtains:

\be
S^{HOM}_{2SC}= e^{2}R(1-R)\sin^{2}(2\tilde{\theta}) \left[1-\cos(\phi_{1}-\phi_{2}) \right]
\label{HOM_2SC}
\ee
which clearly shows a non-local dependence on the difference of the SC order parameter phases as already pointed out in Ref. \onlinecite{Beenakker14}, for the fixed bias case where a continuous current flows. Comparing with Eq. (\ref{current}) we observe that the presented device shows no dependence on the SC phase at the level of the averaged current (first order coherence), but presents an oscillatory modulation in the noise (second order coherence). This purely second order correlation effect is in analogy to what is observed in interferometric geometries discussed in the framework of the quantum Hall effect like the Samuelsson-Sukhorukov-B\"uttiker interferometer \cite{Samuelsson04, Neder07} or the revisitation of the Franson interferometer \cite{Franson89} proposed by Splettstoesser \emph{et al.} \cite{Splettstoesser09} 

It is easy to note that, once the SC phase difference in Eq. (\ref{HOM_2SC}) is different from zero (mod. $2\pi$), the noise vanishes only when two electrons or two holes reach the QPC at the same time ($\tilde{\theta}=0$ and $\tilde{\theta}=\pi/2$ respectively) as a consequence of the Pauli principle. \cite{Jonckheere12} Remarkably enough the noise reaches its maximum for $\tilde{\theta}= \pi/4$. This can be explained in terms of electron/electron and hole/hole interferences occurring at the QPC. In order to better understand this fact, it is useful to take a closer look at the structure of the $\Delta S^{HOM}$ terms in Eq. (\ref{HOM_general}). By considering two Bogoliubov excitations of the form in Eq. (\ref{Bogoliubov}) simultaneously reaching the QPC, it is easy to note that this contribution to the noise is proportional to: 
\beq
&&|\mathcal{W}^{1}_{e} {\mathcal{W}^{2}_{e}}^{*}-\mathcal{W}^{1}_{h} {\mathcal{W}^{2}_{h}}^{*}|^{2}=\nonumber\\
&& | \cos{\tilde{\theta}_{1}} \cos{\tilde{\theta}_{2}}-\sin{\tilde{\theta}_{1}} \sin{\tilde{\theta}_{2}} e^{-i( \Phi_{1}-\Phi_{2})}|^{2}.
\eeq 
It corresponds to the difference between the product of electron and hole probability amplitudes. In particular for $\tilde{\theta}_{1}=\tilde{\theta}_{2}=\pi/4$ the Bogoliubov quasiparticles carry zero charge and zero shot noise, but are by far not trivial excitations with a complex structure given by the coherent superposition of electrons and holes which can be detected only at the level of the two quasiparticle interferometry. The peculiar structure of the HOM noise contribution for two synchronized Bogoliubov excitations directly reflects into the divergences associated with the ratio:

\be
\mathcal{R}_{2SC}= \frac{S^{HOM}_{2SC}}{S^{HBT}_{1}+S^{HBT}_{2}}= -\frac{1}{2} \tan^{2}(2 \tilde{\theta})  \left[1-\cos(\phi_{1}-\phi_{2}) \right].
\ee

\begin{figure}[h]
\centering
\includegraphics[scale=0.4]{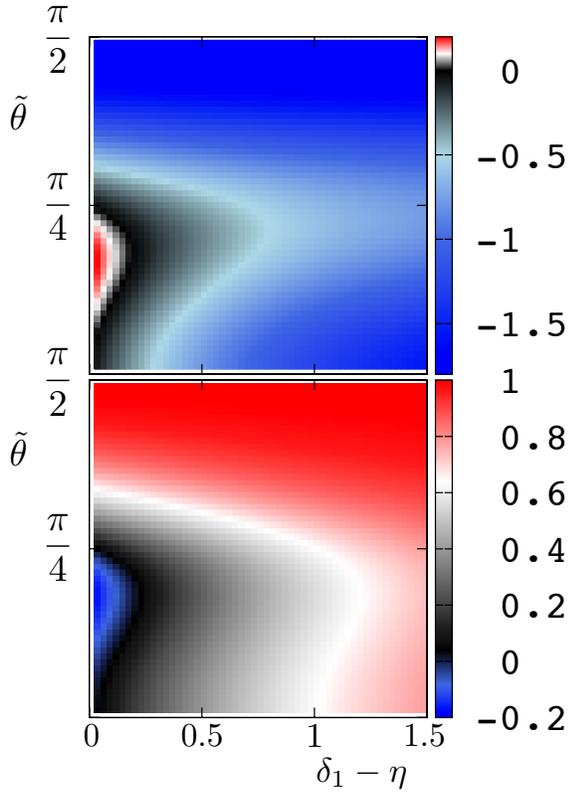}
\caption{(Color online). Upper panel. Density plot of $S^{HOM}_{1SC}$ in units of $S_{0}=e^{2}R(1-R)$. Different shades of blue identify negative regions where the hole contribution dominates over the electron one, while the little red area represents the positive noise. Lower panel. Density plot of $\mathcal{R}_{1SC}$ as a function of $\delta_{1}-\eta$ and $\tilde{\theta}$. The blue area corresponds to negative values of the ratio which cannot be reached in the standard electronic quantum optics experiments involving only electrons and holes at zero temperature.}
\label{fig3}
\end{figure}

We can also achieve collisions between electrons and Bogoliubov quasiparticles. Indeed, when one of the two SC regions ($SES2$ in order to fix the notation) is removed, its transfer matrix reduces to $\mathcal{M}_{2}(\xi)= I\otimes I$ and an injected electron propagates undisturbed along the edge channel until it reaches the QPC. Starting from Eq. (\ref{HOM_general}) one has:

\beq
S^{HOM}_{1SC}&=& e^{2}R(1-R)\times \nonumber \\
&&\left\{\left[ 1+\cos(2\tilde{\theta})\right]A(\delta_{1}-\eta)- \cos^{2}(2\tilde{\theta})-1\right\}.\nonumber\\
\eeq
Here, when the delay is tuned in such a way to have a maximum wave-packet overlap ($\delta_{1}=\eta$ and consequently $A=1$), we can have the reference electron interfering with: a) another electron ($\tilde{\theta}=0$) leading to a zero noise (consequence of the Pauli principle); b) with a hole ($\tilde{\theta}=\pi/2$) with a consequent minimum of the noise\cite{Jonckheere12}; c) a more general Bogoliubov quasiparticle. In the latter case the cross-correlated noise assumes positive values when the electron component of the Bogoliubov excitation dominates over the hole one ($P_{e}>P_{h}$ and consequently for $0<\tilde{\theta}<\pi/4$). By decreasing the wave-packet overlap, the $\Delta S^{HOM}$ contribution to the noise is progressively suppressed. However, even away from perfect synchronization, it is possible to observe positive and negative regions from which we can deduce the dominant contribution to the Bogoliubov quasiparticle. The situation in the case of a wave-packet exponential in time,\cite{Feve07, Ferraro13} where:
\be
A(\tau)= e^{-\Gamma |\tau|}
\ee
with $\Gamma^{-1}$ the escape time of the wave-packet,\cite{Nigg08} is illustrated by the density plot in the upper panel of Fig. \ref{fig3}.   

This represents an extremely useful tool in order to extract information about the structure of the Bogoliubov excitations through interferometric experiments with a known source (electronic wave-packet). These observations indicate that the considered setup offers richer possibilities to implement a tomographic protocol by means of HOM interferometry with respect to what was proposed in the electronic case. \cite{Ferraro14}

Another relevant quantity to explore is given by the ratio: 

\beq
\mathcal{R}_{1SC}&=&  S_{1SC}^{HOM}/(S_1^{HBT}+S_2^{HBT})\nonumber\\
 &=&1- \frac{1+ \cos(2 \tilde{\theta})}{1+ \cos^{2}(2 \tilde{\theta})}A(\delta_{1}-\eta) .
\eeq
This also contains negative regions (blue areas in the lower panel of Fig. \ref{fig3}) which are forbidden in the standard electron quantum optics case ($\tilde{\theta}=0$) \cite{Jonckheere12} due to the constraints imposed by the charge conservation. 


\section{Considerations about possible spin-triplet pairing}\label{Triplet}

\begin{figure}[h]
\centering
\includegraphics[scale=0.5]{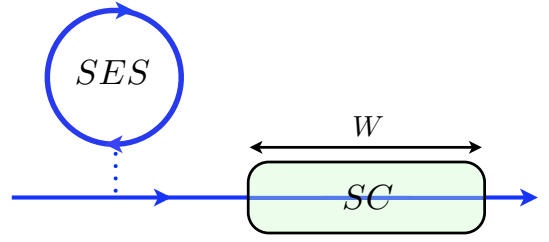}
\caption{(color online). Schematic view of a SES injecting electron and hole wave-packets into a quantum Hall edge state at filling factor $\nu=1$ coupled with a SC contact of length $W$.}
\label{fig4}
\end{figure}

As recently discussed in Ref. \onlinecite{Ostaay11}, an ordinary spin-singlet SC coupling, together with a strong Rashba spin-orbit interaction in single quantum Hall edge channel can lead to an effective spin-triplet coupling and to a consequently small, but not negligible effect of Andreev reflections at the interface between the edge state and the SC region. This effect is predicted to be present in heterostructures based of InAs and InSb, where the measurement of magneto-resistance for systems in the Hall regime and in the presence of SC has been recently carried out. \cite{Eroms05, Batov07} In order to compare this case with the previous one we will consider a fully polarized chiral quantum Hall edge channel (at filling factor $\nu=1$) coupled to a SC contact (see Fig. \ref{fig4}). Under these conditions the spin degree of freedom can be neglected. 

The action of the SC in this case can be described in terms of a $2\times 2$ transfer matrix \cite{Ostaay11, Beenakker14} constrained again by unitarity and particle-hole symmetry \cite{Note3}
\be
\mathcal{M}(\xi)= \tau_{x} \mathcal{M}^{*}(-\xi) \tau_{x}.
\label{particle_hole2}
\ee

It naturally acts on the $2$-component spinor state 

\be
c(\xi)= 
\left( 
\begin{matrix}
c_{e}(\xi) \\
c_{h}(\xi)\\
\end{matrix}
\right)
\label{spinor}
\ee
which, according to Eq. (\ref{particle_hole2}), satisfies \cite{Alicea12}
 
 \be
 c(\xi)= \tau_{x} \left[ c^{\dagger}(-\xi)\right]^{T}.
 \ee
 
The transfer matrix can be written formally as in Eq. (\ref{transfer}), with 

\be
\mathcal{U}(\alpha, \phi, \beta) =\left( 
\begin{matrix}
\cos{\theta}+i \frac{\beta}{\theta}\sin{\theta} & i \frac{\alpha}{\theta} e^{-i \phi} \sin{\theta}\\
 i \frac{\alpha}{\theta} e^{i \phi} \sin{\theta}&\cos{\theta}-i \frac{\beta}{\theta}\sin{\theta}\\
\end{matrix}
\right)\label{U_matrix}
\ee
which is rewritten in terms of the $\tilde{\theta}$ angle: 
\be
\mathcal{U}(\tilde{\theta}, \phi, 0) =\left( 
\begin{matrix}
\cos{\tilde{\theta}}e^{i \Omega} & i e^{-i \phi} \sin{\tilde{\theta}}\\
 i e^{i \phi} \sin{\tilde{\theta}}&\cos{\tilde{\theta}}e^{-i \Omega}\\
\end{matrix}
\right).
\ee
Nevertheless the microscopic derivation of the parameters starting from the Bogoliubov-De Gennes equation is different. \cite{Ostaay11} In order to better understand the analogies and differences with respect to the spin-singlet case, it is useful to consider two parameters, namely the velocity of propagation along the quantum Hall edge channel: 

\be
v\approx \omega_{c} l_{m},
\ee
where $\omega_{c}$ is the cyclotron frequency and: 
\be
v_{\Delta}\approx v \frac{d}{l_{so}}
\ee
with $d$ the length characterizing the variations of the electrostatic potential at the interface between the SC and the Hall channel and: 
\be
l_{so}= \frac{\hbar^{2}}{m a}
\ee
the Rashba length ($m$ the effective mass of the electrons, $a$ the Rashba spin-orbit coefficient). Typically in experiments one has $v\gg v_{\Delta}$. According to Ref. \onlinecite{Ostaay11}, the time required to cross the SC region is given by: 

\be
\delta= W \frac{v}{v^{2}-v^{2}_{\Delta}}\approx \frac{W}{v}
\ee
in analogy to what is observed for the spin-singlet pairing case. The same holds also for the parameter: 

\be
\beta\approx \frac{W}{l_{m}}.
\ee
The important difference concerns the parameter $\alpha$ which is energy dependent and is given by: 

\be
\alpha(\xi) = \xi W \frac{v_{\Delta}}{v^{2}-v^{2}_{\Delta}}\approx \xi W \frac{v_{\Delta}}{v^{2}}\approx W \frac{\xi}{v} \frac{d}{l_{so}}.
\ee
The above relation clearly show that $\alpha(\xi)=-\alpha(-\xi)$. This requires some additional comments about the nature of the experimentally realizable electronic wave-packet. On one hand the so called Levitons, electronic wave-packets obtained by applying a properly quantized Lorentzian voltage in time \cite{Dubois13, Grenier13, Dubois13b} are intrinsically emitted near the Fermi level \cite{Ferraro13, Jullien14} (close to zero energy), where $\alpha \approx 0$ and therefore the SC device has no effect. On the other hand the  SES described in Ref. \onlinecite{Feve07} emits, in the optimal regime, a wave-packet with a well defined energy (frequency) $\omega_{0}$ above the Fermi sea. Here, we can safely approximate $\alpha$ as a constant given by $\alpha(\omega_{0})$. In this case one finds that 

\be
\frac{\beta}{\alpha{(\omega_{0})}}\approx \frac{\omega_{c}}{\omega_{0}}\frac{l_{so}}{d}\gg 1,
\label{k2}
\ee
where $\hbar \omega_{0}$ is constrained by the induced SC gap $\Delta$ (few Kelvin) and $\hbar \omega_{c}$ the energy gap of the Hall fluid (around ten Kelvin in the integer regime). 

We are therefore in the opposite regime with respect to what we discussed for the spin-singlet case. Although in the present case we can introduce a projector: 

\be
\mathcal{P}_{t}= \left(\frac{1+\tau_{z}}{2}\right)
\ee
which represents the injection of a purely electronic wave-packet in order to obtain \emph{exactly the same formulas as before} for all the transport properties (current, noise, HBT and HOM correlators),  one realizes that the condition in Eq. (\ref{k2}) associated to the defnition in Eq. (\ref{sin_theta}) forces: 
\be
\sin{\tilde{\theta}}\ll 1 
\ee
leading only to small SC corrections to the physics of the SES. 

This suggests that the spin-triplet coupling is not optimal in order to realize interferometric experiments involving individual  Bogoliubov quasiparticles.

\section{Conclusions}\label{Conclusions}

This paper was devoted to the study of HBT and HOM interferometry of single Bogoliubov quasiparticles, which are potential candidates for Majorana fermions.
We started with the description of the single quasiparticle source. The setup is composed of a quantum Hall edge channel coupled with a SES and a SC contact. This device behaves as an emitter of individual Bogoliubov excitations, namely coherent superpositions of electrons and holes. The current outgoing from it is given by the incoming one -- albeit delayed in time and multiplied by a prefactor reminiscent of the action of the Andreev reflection -- while the excitation density is simply delayed due to the conservation of the excitation number. 
The zero frequency noise associated with the source depends on the mixing angle which controls the electron and hole content of the quasiparticle wave packet. In particular, we obtain a generalization of  the results for zero frequency noise observed for the SES in absence of SC contact. 

The controlled emission of Bogoliubov excitations can be used to realize electron quantum optics experiments such as HBT and HOM interferometry. In the former (HBT) case we obtain the shot noise associated with the (non-integer) charge of the Bogoliubov excitation. It vanishes when the electron component of this object corresponds to its hole counterpart. 
In the latter (HOM) it is possible to investigate two particle interference properties of these peculiar excitations showing that quasiparticles bearing zero charge and thus zero partition noise have maximal HOM contribution as a consequence of interference between the electron/electron and hole/hole amplitudes.
This is explicit in our zero time delay predictions. 

We finally proposed a source injecting purely electronic wave-packets as a way to realize the HOM spectroscopy of the Bogoliubov excitations. Plots of the HOM noise revealed that this quantity can either be positive or negative depending on the weight of the electron or hole component of the Bogoliubov quasiparticle. This result has no equivalent in current electron optics experiments.  

For completeness, we also considered the triplet pairing case for a single edge state, where our formalism can be translated straighforwardely from the $\nu=2$ case. However, our estimates for experimental investigations in this former case suggest that the mixing angle is confined to low values, which does not constitute an optimal setup for the observation of non local interference phenomena of single Bogoliubov quasiparticles.
Our analysis clearly showed that the singlet-spin coupling at filling factor $\nu=2$ is more suitable for that respect.

Extensions of this work could include the discussion of finite temperature effects, as can be readily achieved in scattering matrix approaches to electronic quantum optics calculations.\cite{Jonckheere12} More demanding would be to include the effect of Coulomb interaction between edges, as the phenomenon of electron fractionalization which is known to occur at $\nu=2$ in the absence of superconductivity, and which gives rise to a charged and a neutral mode, would modify the nature of Bogoliubov quasiparticles. 

Concerning the experimental feasibility, we point out that on demand electron sources are currently available,\cite{Feve07,Dubois13b} and that the conditions for placing a superconductor in contact to quantum Hall edge channels have been discussed previously.\cite{Beenakker14} Single particle sources typically operate with periodic trains of electrons and holes, implying many collisions between the injected objects, which allow for data acquisition, because a single shot experiment with two particles (quasiparticles) cannot be achieved so far. In addition to the spectroscopy experiment proposed in this work with electrons and Bogoliubov quasiparticles, we can envision collisions between holes and Bogoliubov quasiparticles, which constitutes a straightforward extention of our present results. An important experimental advantage with the present single particle sources proposal is that it allows to probe the description of the annihilation of Bogoliubov particle (thus Majorana fermions) at the single excitation level, and this protocol calls ``only'' for zero frequency noise measurements, as opposed to the biased voltage/``quasiparticle beam'' experiments suggested in Ref. \onlinecite{Beenakker14}. Thus the experiments proposed here do no require an on-chip noise measurement circuit in order to achieve the diagnosis of non local interference and (in  particular) the annihilation of single Bogoliubov quasiparticles.    

\acknowledgements

The authors acknowledge the support of Grant No. ANR-2010-BLANC-0412 (``1 shot'') and of ANR-2014-BLANC ``one shot reloaded''. This work has been carried out
in the framework of the Labex ARCHIMEDE (Grant No. ANR-11-LABX-0033) and of the A*MIDEX project (Grant No. ANR-
11-IDEX-0001-02), funded by the ``investissements d'avenir'' French Government program managed by the French National
Research Agency (ANR).

\appendix

\section{Current and noise of the source} \label{AppA}
In this Appendix we evaluate explicitly the main quantities needed in order to fully characterize the behavior of the source described in Fig. \ref{fig1}. As achieved in the main text we consider the injection of a purely electronic wave-packet by the SES. 

\subsection{Averaged current}
According to Eq. (\ref{current_def}), the averaged outgoing current is given by: 

\be
\langle \varphi |I(t) |\varphi \rangle =-e v\langle \varphi |:\Psi^{\dagger}(t) \tau_{z} \Psi(t): |\varphi\rangle.
\ee
By properly considering an incoming spin up electron in the energy domain (see Eq. (\ref{wave_packet}))
\be
|\varphi \rangle =\frac{1}{\sqrt{2 \pi}} \int^{+\infty}_{-\infty} d \xi \tilde{\varphi}_{e}(\xi) c^{\dagger}_{e, \uparrow}(\xi) |F\rangle,
\label{wave}
\ee
and the field operator outgoing from the SC region 
\be
\Psi(t)=\frac{1}{\sqrt{4\pi}} \int^{+\infty}_{-\infty} d \xi e^{-i \xi t} \mathcal{M}(\xi) c(\xi) 
\label{field}
\ee
 the expression for the current becomes: 
 \beq
 \langle \varphi |I(t) |\varphi \rangle &=&-\frac{e v}{4 \pi} \int^{+\infty}_{-\infty} d \xi d \eta e^{i \eta t} e^{- i \xi t} \nonumber \\
 &&\langle \varphi |: c^{\dagger}(\eta) \mathcal{M}^{\dagger}(\eta) \tau_{z} \mathcal{M}(\xi) c(\xi): |\varphi\rangle.
 \eeq
Due to the fact that we are dealing with non-interacting electrons, we can safely apply Wick's theorem in order to evaluate the correlation functions. Moreover, it is useful to exploit the particle-hole symmetry of the system in such a way to constraint the integrals only in the positive energy interval and to avoid problems related to double-counting. Once considered the incoming wave-packet, the above expression reduces to: 

 \beq
 \langle \varphi |I(t) |\varphi \rangle &=&-\frac{e}{(2 \pi)^2} \mathrm{Tr} \left( \mathcal{P}_{s} \tilde{\mathcal{M}}^{\dagger} \tau_{z} \tilde{\mathcal{M}} \mathcal{P}_{s}\right) \times\nonumber\\
 && \int^{+\infty}_{0} d \xi d \eta e^{i \eta (t-\delta)} e^{- i \xi (t-\delta)} \tilde{\varphi}^{*}(\eta) \tilde{\varphi}(\xi)\nonumber\\
\eeq
where we have considered the definition in Eq. (\ref{transfer}) and the projector over the the electronic state with spin up $\mathcal{P}_{s}$ (see Eq. (\ref{projector})). 

For electronic wave-packets with energy components only above the Fermi level we can safely extend again the domain of integration in the interval ($-\infty$, $+\infty$) in order to deal with the complete Fourier transform of the wave-packets, obtaining 
\beq
 \langle \varphi |I(t) |\varphi \rangle &=&-e \mathrm{Tr} \left( \mathcal{P}_{s} \tilde{\mathcal{M}}^{\dagger} \tau_{z} \tilde{\mathcal{M}} \mathcal{P}_{s}\right) \varphi^{*}(t-\delta) \varphi(t-\delta) \nonumber\\
 \\
 &=&-e \cos(2 \tilde{\theta}) \varphi^{*}(t-\delta) \varphi(t-\delta) 
 \eeq
where in the last line we have considered the explicit form of the transfer matrix from Eq. (\ref{transfer_exp}).

\subsection{Particle density}

For the particle density introduced in Eq. (\ref{density})
\be
\langle \varphi |\rho (t) |\varphi \rangle = v\langle \varphi |:\Psi^{\dagger}(t) \Psi(t): |\varphi \rangle
\ee
one can proceed along the same lines and obtains: 

\beq
\langle \varphi |\rho (t) |\varphi \rangle &=&\mathrm{Tr} \left( \mathcal{P}_{s} \tilde{\mathcal{M}}^{\dagger} I \tilde{\mathcal{M}} \mathcal{P}_{s}\right) \varphi^{*}(t-\delta) \varphi(t-\delta)\nonumber\\
&=& \varphi^{*}(t-\delta) \varphi(t-\delta) 
\eeq
where, in this case, the trace reduces to one due to the unitarity of the transfer matrix $\mathcal{M}$, ultimately leading to a result compatible with the particle number conservation. 

\subsection{Noise}
The noise outgoing from the source is defined, in the wave-packet limit, as (see Eq. (\ref{noise_out}))

\beq
S_{source}&=& \int^{+\infty}_{-\infty} dt dt' \left[\langle \varphi| I(t) I(t')  |\varphi\rangle- \langle \varphi| I(t)|\varphi \rangle \langle \varphi | I(t')  |\varphi\rangle\right] \nonumber\\
&=&e^{2} v^{2} \int^{+\infty}_{-\infty} dt dt' \left[\langle \varphi| \Psi^{\dagger}(t) \tau_{z} \Psi(t)  \Psi^{\dagger}(t') \tau_{z} \Psi(t')|\varphi\rangle-\right. \nonumber\\ 
&&\left. \langle \varphi| \Psi^{\dagger}(t) \tau_{z} \Psi(t)|\varphi\rangle  \langle \varphi| \Psi^{\dagger}(t') \tau_{z} \Psi(t')|\varphi\rangle\right].
\eeq
By replacing the expressions for the field operator in Eq. (\ref{field}) and the incoming wave-packet in Eq. (\ref{wave}) it is possible to develop the calculation in full analogy with what was done before. The first contribution to the noise is given by 
\beq
&&e^{2} v^{2} \int^{+\infty}_{-\infty} dt dt' \langle \varphi| \Psi^{\dagger}(t) \tau_{z} \Psi(t)  \Psi^{\dagger}(t') \tau_{z} \Psi(t')|\varphi\rangle= \nonumber\\
&& e^{2}\mathrm{Tr}\left(\mathcal{P}_{s} \tilde{\mathcal{M}}^{\dagger} \tau_{z} \tilde{\mathcal{M}}\tilde{\mathcal{M}}^{\dagger} \tau_{z}\tilde{\mathcal{M}}\mathcal{P}_{s}\right)= e^{2} \mathrm{Tr}\left(\mathcal{P}_{s}\right)=e^{2}, \nonumber\\
\eeq
where we have considered the unitarity of the transfer matrix $\tilde{\mathcal{M}}$ and of the Pauli matrix $\tau_{z}$.

The second contribution reads: 

\beq
&&e^{2} v^{2}\int^{+\infty}_{-\infty} dt dt' \langle \varphi| \Psi^{\dagger}(t) \tau_{z} \Psi(t)|\varphi\rangle  \langle \varphi| \Psi^{\dagger}(t') \tau_{z} \Psi(t')|\varphi\rangle= \nonumber\\
&&e^{2} \left[ \mathrm{Tr} \left( \mathcal{P}_{s} \tilde{\mathcal{M}}^{\dagger} \tau_{z} \tilde{\mathcal{M}} \mathcal{P}_{s}\right)\right]^{2}= e^{2} \cos^{2}(2 \tilde{\theta})
\eeq
in full agreement with what was obtained for the current.

Recollecting all the above results we finally obtain 

\be
S_{source}= e^{2} \left[1-  \cos^{2}(2 \tilde{\theta})\right]= e^{2} \sin^{2}(2 \tilde{\theta}).
\ee

\subsection{HBT and HOM signal}

Proceeding in the same way as before, the cross-correlation of the currents outgoing from the QPC in the geometry of Fig. \ref{fig2} is given by
\begin{widetext}
\beq
S&=& e^{2} v^{2} \int^{+\infty}_{0} d \xi d \rho \left[\langle \phi| a^{\dagger}_{1}(\xi) \tau_{z} a_{1}(\xi)  a^{\dagger}_{2}(\rho) \tau_{z} a_{2}(\rho) |\phi\rangle-\langle \phi| a^{\dagger}_{1}(\xi) \tau_{z} a_{1}(\xi)|\phi\rangle \langle \phi |  a^{\dagger}_{2}(\rho) \tau_{z} a_{2}(\rho) |\phi\rangle\right]\nonumber\\
&=&  e^{2} v^{2} R(1-R) \int^{+\infty}_{0} d \xi d \rho \left[\langle \phi | c^{\dagger}_{1}(\xi) \mathcal{M}^{\dagger}_{1}(\xi) \tau_{z}\mathcal{M}_{1}(\xi) c_{1}(\xi) c^{\dagger}_{1}(\rho)\mathcal{M}^{\dagger}_{1}(\rho) \tau_{z} \mathcal{M}_{1}(\rho) c_{1}(\rho) \right.\nonumber \\
&-& \langle \phi | c^{\dagger}_{1}(\xi) \mathcal{M}^{\dagger}_{1}(\xi) \tau_{z}\mathcal{M}_{2}(\xi) c_{2}(\xi) c^{\dagger}_{2}(\rho)\mathcal{M}^{\dagger}_{2}(\rho) \tau_{z} \mathcal{M}_{1}(\rho) c_{1}(\rho) \nonumber \\
&-& \langle \phi | c^{\dagger}_{2}(\xi) \mathcal{M}^{\dagger}_{2}(\xi) \tau_{z}\mathcal{M}_{1}(\xi) c_{1}(\xi) c^{\dagger}_{1}(\rho)\mathcal{M}^{\dagger}_{1}(\rho) \tau_{z} \mathcal{M}_{2}(\rho) c_{2}(\rho) \nonumber \\
&+& \langle \phi | c^{\dagger}_{2}(\xi) \mathcal{M}^{\dagger}_{2}(\xi) \tau_{z}\mathcal{M}_{2}(\xi) c_{2}(\xi) c^{\dagger}_{2}(\rho)\mathcal{M}^{\dagger}_{2}(\rho) \tau_{z} \mathcal{M}_{2}(\rho) c_{2}(\rho) \nonumber \\
&-&\langle \phi | c^{\dagger}_{1}(\xi) \mathcal{M}^{\dagger}_{1}(\xi) \tau_{z}\mathcal{M}_{1}(\xi) c_{1}(\xi)|\phi \rangle \langle \phi | c^{\dagger}_{1}(\rho)\mathcal{M}^{\dagger}_{1}(\rho) \tau_{z} \mathcal{M}_{1}(\rho) c_{1}(\rho) \nonumber \\
&-&\left. \langle \phi | c^{\dagger}_{2}(\xi) \mathcal{M}^{\dagger}_{2}(\xi) \tau_{z}\mathcal{M}_{2}(\xi) c_{2}(\xi)|\phi \rangle \langle \phi | c^{\dagger}_{2}(\rho)\mathcal{M}^{\dagger}_{2}(\rho) \tau_{z} \mathcal{M}_{2}(\rho) c_{2}(\rho)\right].
\eeq
\end{widetext}
According to this the HOM contribution to the noise can be written as 
\begin{widetext}
\beq
S^{HOM}&=& e^{2} R (1-R)\left\{2 \mathrm{Tr} \left[ \mathcal{P}_{s} \tilde{\mathcal{M}}_{1}^{\dagger}\tau_{z}  \tilde{\mathcal{M}}_{2} \mathcal{P}_{s}\right] \mathrm{Tr} \left[ \mathcal{P}_{s} \tilde{\mathcal{M}}_{2}^{\dagger}\tau_{z}  \tilde{\mathcal{M}}_{1} \mathcal{P}_{s}\right] A(\delta_{1}- \delta_{2}-\tau_{1}+\tau_{2}) \nonumber \right.\nonumber \\
&-&\left.\mathrm{Tr} ^{2}\left[ \mathcal{P}_{s} \tilde{\mathcal{M}}_{1}^{\dagger}\tau_{z}  \tilde{\mathcal{M}}_{1} \mathcal{P}_{s}\right]- \mathrm{Tr} ^{2}\left[ \mathcal{P}_{s} \tilde{\mathcal{M}}_{2}^{\dagger}\tau_{z}  \tilde{\mathcal{M}}_{2} \mathcal{P}_{s}\right]\right\}\nonumber\\
&=& e^{2}R(1-R)\left\{\left[ 1+ \cos(2\tilde{\theta}_{1})\cos(2\tilde{\theta}_{2})-\cos(\Phi_{12}) \sin(2\tilde{\theta}_{1})\sin(2 \tilde{\theta}_{2})\right]A(\delta_{1}-\delta_{2}-\tau_{1}+\tau_{2}) \right.\nonumber\\ 
&-&\left. \cos^{2}{(2\tilde{\theta}_{1})}-\cos^{2}{(2\tilde{\theta}_{2})} \right\}.
\eeq
\end{widetext}
while the HBT contribution is simply obtained imposing the condition $\mathcal{M}_{j}=I \otimes I$ to one of the two transfer matrices in the above expression.

\end{document}